# Enhanced Tumor Accumulation of Sub-2 nm Gold Nanoclusters for Cancer Radiation Therapy

*Xiao-Dong Zhang[1], Jie Chen[1], Zhentao Luo[2], Di Wu[1], Xiu Shen[1], Sha-Sha Song[1], Yuan-Ming Sun[1], Pei-Xun Liu[1], Jing Zhao[3], Shuaidong Huo[3], Saijun Fan[1], Feiyue Fan[1], Xing-Jie Liang[3]\*, and Jianping Xie [3]\**

[1]Tianjin Key Laboratory of Molecular Nuclear Medicine, Institute of Radiation Medicine, Chinese Academy of Medical Sciences and Peking Union Medical College, Tianjin, 300192, China

[2]Department of Chemical and Biomolecular Engineering, National University of Singapore, 10 Kent Ridge Crescent, 119260, Singapore

[3]CAS Key Laboratory for Biological Effects of Nanomaterials and Nanosafety, National Center for Nanoscience and Technology, Beijing, 100190, China

Email: chexiej@nus.edu.sg
Email: liangxj@nanoctr.cn

**Abstract**: A new type of *metabolizable* and *efficient* radiosensitizer for cancer radiotherapy is presented in this study by combining ultrasmall Au nanoclusters (NCs, <2 nm) with biocompatible coating ligands (glutathione, GSH). The new nano-construct (GSH-coated $Au_{25}$ NCs) inherits attractive features of both the Au core (strong radiosensitizing effect) and GSH shell (good biocompatibility). It can preferentially accumulate in tumor *via* the improved EPR effect, which leads to strong enhancement for cancer radiotherapy. After the treatment, the small-sized GSH-$Au_{25}$ NCs can be efficiently cleared by the kidney, minimizing any potential side effects due to the accumulation of $Au_{25}$ NCs in the body.

## 1. Introduction

Gold nanoparticles (Au NPs) possess distinct physical and chemical properties that make them ideal platforms for a variety of biomedical applications, including imaging, biosensing, drug delivery, and therapy.[1] Recently, Au NPs have been applied as a promising new type of radiotherapy sensitizing agents (or radiosensitizers) for cancer treatment because of their strong absorption and high efficiency in generating secondary electrons under gamma ray or X-ray irradiation, enhancing the DNA and protein damages in tumors.[2] A good radiosensitizer must feature two properties: (1) efficient accumulation in tumors during the radiation treatment to achieve sufficient enhancement for the radiotherapy, and (2) effective renal clearance after the treatment to minimize toxic side effects. Here we report a novel type of Au NP-based radiosensitizers: ultrasmall Au NPs (<2 nm) protected by naturally occurring biomolecules (e.g., peptides and proteins). The complementary features of the two key components, the sub-2 nm Au core and the biomolecule coating, are integrated into a single entity, that is, biomolecule-coated ultrasmall Au NPs, to achieve targeted properties of a good radiosensitizer.

Ultrasmall Au NPs were chosen in this study as the radiosensitizer. Sub-2 nm Au NPs, or termed Au nanoclusters (NCs), are a subgenre of NPs with a core size below 2 nm typically containing less than 150 Au atoms.[3] Recent studies have shown that the accumulation efficiency of Au NPs in tumors was largely determined by the particle size, and smaller Au NPs have higher efficient tumor deposition. For example, it is well-documented that Au NPs with sizes >50 nm cannot pass through the outside barrier of the reticuloendothelial system (RES) but to form large aggregates (>100 nm) during the blood circulation, which led to a poor deposition of NPs in tumors.[4] In contrast, Au NPs with sizes <50 nm showed radiation enhancement effects to a certain extent, suggesting the accumulation of the NPs in tumors.[2a, 5] In addition, 12 nm Au NPs coated with polyethylene glycol (PEG) showed obviously higher accumulation in tumors than those of 27.6 and 46.6 nm PEG-coated Au NPs.[2b, 6] It is therefore expected that the ultrasmall Au NCs (< 2 nm) may have improved accumulation in tumors because of their enhanced permeability and retention (EPR) effect relative to that of large NPs (>2 nm),[7] leading to an improved enhancement for cancer radiotherapy.

Coating ligands on the NP surface can also affect the pharmacokinetics of Au NPs in the body by controlling the surface chemistry and hydrodynamic diameter (HD) of the Au NPs, which are two primary factors that can affect the properties and behavior of Au NPs under physiological conditions.[8] For example, 3 nm Au NPs coated with different ligands showed different colloidal stability in the body: the PEG-coated Au NPs were not stable and formed aggregates in blood, whereas the glutathione (GSH)-coated Au NPs were highly stable under the same conditions.[9] The different behaviors of Au NPs in blood may lead to different deposition efficiencies in tumors. Due to the ultrasmall size of Au NCs, the effect of coating ligands on their pharmacokinetics in the body could be more pronounced than that of larger NPs.[10] In addition, the biocompatibility of the coating ligands on the NP surface is a key consideration for all biomedical applications, and a good compliance strategy is to select a naturally occurring biomolecule as the coating ligand.[11] The natural peptide – GSH and the bovine serum albumin (BSA) are widely used as biocompatible coating ligands for Au NPs in various biomedical settings. Therefore, we hypothesized that an efficient and safe radiosensitizer could be constructed by incorporating biocompatible ligands (GSH or BSA) into the ultrasmall Au NCs, forming GSH- or BSA-coated Au NCs, which may have improved passive tumor targeting performance *via* the EPR effect.

One concern related to the therapeutic Au NPs is biosafety. Au NPs are generally considered as safe materials in view of their good biocompatibility and low *in vitro* cytotoxicity.

However, Au NPs with a relatively large size (>10 nm) cannot be metabolized and tend to be absorbed by RES and accumulate in liver and spleen, which leads to potential damages to the liver and immune system.[12] Recent studies suggested that the HD of the NPs could determine the renal clearance efficiency.[9a] For example, semiconductor quantum dots (QDs) with a HD <5.5 nm was rapidly and efficiently removed by the kidney, whereas QDs with a HD >15 nm did not show efficient renal excretion and were accumulated in the liver and spleen.[13] Similarly, the GSH-coated Au NPs with particle size in the range of 1.5–3 nm showed high efficiency in the renal clearance.[9-10]

This article is an account of our investigation of the GSH- and BSA-coated Au NCs for cancer radiotherapy. The well-studied $Au_{25}$ NCs species was chosen as the model of the new NC-based radiosensitizers because of their ultrasmall size (<2 nm) and excellent chemical stability. *In vitro* and *in vivo* studies of the GSH- and BSA-coated $Au_{25}$ NCs (or GSH- and BSA-$Au_{25}$ NCs for short) for cancer radiotherapy are presented here with detailed analyses of the cell response, DNA damage, and changes in tumor volume and weight after the treatment. *In vivo* cytotoxicity of GSH-$Au_{25}$ NCs was investigated with experimental evidences from the pathology, biochemistry, organ index, and biodistribution studies. Our findings suggest that GSH-$Au_{25}$ NCs are promising radiosensitizers for cancer radiotherapy with attractive features including excellent tumor accumulation, strong radiation enhancement, and low toxicity (efficient renal clearance).

## 2. Results and Discussion

GSH- and BSA-Au$_{25}$ NCs were synthesized and purified according to published procedures.[14] Both types of the NCs have a core-shell structure, as illustrated in **Figure 1**a and 1b. GSH-Au$_{25}$ NCs, which can be denoted as Au$_{25}$(SG)$_{18}$, have molecular-like absorptions due to the strong quantum confinement of free electrons in the ultrasmall particles.[15] As shown in Figure 1c (black line), the GSH-Au$_{25}$ NCs showed the UV-vis absorption spectrum with a maximum at 670 nm and a few shoulder peaks that matched well with earlier studies.[3d, 16] The well-defined absorption spectrum suggested the high purity of our sample. The ultrasmall size of GSH-Au$_{25}$ NCs was demonstrated by the transmission electron microscopy (TEM) image (Figure 1e), in which only particles smaller than 1.5 nm were observed. BSA-Au$_{25}$ NC consists of 25 Au atoms encapsulated by a BSA molecule (Figure 1b). Unlike GSH-Au$_{25}$ NCs, BSA-Au$_{25}$ NCs showed no obvious peaks in its UV-vis absorption spectrum (Figure 1c, red line). The distinctive difference in the absorptions of GSH- and BSA-Au$_{25}$ NCs is a result of the strong effects from the different coating ligands and protection chemistries, which are more pronounced in the ultrasmall NCs than in large NPs.[14a, 17] As suggested by the TEM image (Figure 1f), the core size of BSA-Au$_{25}$ NCs was also below 1.5 nm, similar to that of GSH-Au$_{25}$ NCs. However, GSH- and BSA-Au$_{25}$ NCs showed very different HDs in dynamic light scattering (DLS) measurements. The HD of GSH-Au$_{25}$ NCs was ~2.4 nm (Figure 1d, black line), which was determined by the small size of the GSH ligand [molecular weight (MW) = 307]. In contrast, the BSA-Au$_{25}$ NCs had a HD of ~6 nm, which matched nicely with the size of BSA (MW ~66 kDa and HD ~6 nm) because the Au atoms were encapsulated by a single BSA molecule for each BSA-Au$_{25}$ NC.

The cell responses (or biocompatibility) of GSH- and BSA-Au$_{25}$ NCs were evaluated using the Hela cells. The Hela cell culture was first treated with GSH- or BSA-Au$_{25}$ NCs at different concentrations of 0.00625–0.2 mg-Au/mL. As shown in **Figure 2**a, after 24 and 48 h, the viability of the cells changed very little with increasing NC concentration. The results showed that both GSH- and BSA-Au$_{25}$ NCs had low cytotoxicity even at a high dosage of 0.2 mg-Au/mL (after 48 h, the cell viability was ~85% and ~70% for GSH- and BSA-Au$_{25}$ NCs, respectively). The data also suggested that GSH-Au$_{25}$ NCs had better biocompatibility than that of BSA-Au$_{25}$ NCs. The good biocompatibility of GSH- and BSA-Au$_{25}$ NCs is expected because both GSH and BSA are naturally occurring benign biomolecules.

The radiation enhancements of GSH- and BSA-Au$_{25}$ NCs were measured by the colony formation assay using the Hela cells. As shown in Figure 2b, an obvious enhancement in radiation was observed for cell cultures treated with the GSH- or BSA-Au$_{25}$ NCs. In particular, the sensitization enhancement ratio (SER) of GSH-Au$_{25}$ NCs was ~1.30, which was higher than that of BSA-Au$_{25}$ NCs (~1.21) for all the radiation doses. The radiation enhancement effects of GSH- and BSA-Au$_{25}$ NCs may be due to the enhanced DNA damage induced by the photoelectric effect and Compton scattering of the heavy metal, that is, Au. This hypothesis was supported by the single-cell gel electrophoresis study. As shown in Figure 2c, without radiation, negligible DNA damage was observed for the cell cultures treated with GSH- and BSA-Au$_{25}$ NCs. This data provided yet another line of evidence for low cytotoxicity of both GSH- and BSA-Au$_{25}$ NCs. In contrast, after receiving a 3 Gy radiation dose, a significant DNA damage was observed for cell cultures treated with GSH- and BSA-Au$_{25}$ NCs (Figure 2c). The remarkable DNA damage was also suggested by the *in vitro* imaging with fluorescent DNA stain, in which a long tail indicated significant DNA damage. There was no obvious DNA damage observed in the control and NCs without radiation groups (Figure S1). As compared with the cell culture treated by radiation only (Figure 2d), cell cultures treated with GSH- or BSA-Au$_{25}$ NCs *plus* radiation (3 Gy) showed more significant DNA damages (Figure 2e and 2f). In addition, GSH-Au$_{25}$ NCs showed stronger radiation enhancement than BSA-Au$_{25}$ NCs, which could be attributed to the improved cell uptake of the hydrodynamically smaller GSH-Au$_{25}$ NCs (HD ~2.4 nm) relative to that of the BSA-Au$_{25}$ NCs (HD ~6 nm). The different surface chemistry of the coating ligands (GSH and BSA) on the NCs might also contribute to the difference in their cell uptake. GSH is a ziwitterionic ligand with two carboxyl and one amine groups, which might enhance the uptake of the GSH-Au$_{25}$ NCs by the cells.

*In vivo* experiments were carried out to further confirm the strong radiation enhancement caused by GSH- or BSA-Au$_{25}$ NCs. We first labeled GSH- and BSA-Au$_{25}$ NCs with an organic dye, Cy5, whose excitation and emission wavelengths were 595 and 680 nm, respectively. The Cy5-labeled GSH- and BSA-Au$_{25}$ NCs also showed red emission peaks at around 680 nm when excited at 595 nm (Figure S2). The *in vivo* pharmacokinetics of GSH- and BSA-Au$_{25}$ NCs in blood were evaluated using male nude mice bearing the U14 tumor with a tumor weight of ~10 mg-tumor/kg-body. The mice were intraperitoneally injected with Cy5-labeled GSH- or BSA-Au$_{25}$ NCs (10 mg-Au/kg-body), and the concentrations of GSH- or BSA-Au$_{25}$ NCs in blood were monitored by their fluorescence over a 24 h period after injection. As shown in **Figure 3**a, the half-life of GSH- and BSA-Au$_{25}$ NCs in blood was determined to be ~0.25 and 0.75 h, respectively. As compared with the BSA-Au$_{25}$ NCs, the shorter half-life of GSH-Au$_{25}$ NCs in blood could be attributed to its much smaller HD. The concentrations of both GSH- and BSA-Au$_{25}$ NCs in blood were gradually stabilized after ~2.5 h (Figure 3a). The pharmacokinetics of such passive targeting NCs may follow a two-compartment model in which the biocompatible ligands on the NC surface help the NCs in blood penetrate tissues through the transendothelial pores in tumor blood vessels and subsequently deposit in the tumor interstitium.[18]

The accumulations of GSH- and BSA-Au$_{25}$ NCs in tumors were further confirmed by the *in vivo* fluorescence imaging (Figure 3b). Strong fluorescence was observed in the tumor site (indicated by the circle) at 24 h after the injection of Cy5-labeled GSH- or BSA-Au$_{25}$ NCs. Strong fluorescence was also observed in the liver and bladder. The *in vivo* biodistributions of the injected GSH- and BSA-Au$_{25}$ NCs were evaluated using the fluorescence intensity of the tissues at 680 nm when excited by a 595-nm laser. Figure 3c shows that the NCs were rarely found in most of the organs except liver and bladder, indicating that the NCs had sufficient transit time in the systemic circulation for the deposition in tumor. The depositions of GSH- and BSA-Au$_{25}$ NCs in tumors were ~13.1% and 8.6% ID/g (refers to percentage of the injected dose per gram-tissue), respectively. The *in vivo* biodistribution data of the NCs at day 20 after injection (Figure S3) was also investigated by measuring the concentrations of Au in the dissected organs of sacrificed mice bearing the U14 tumors using inductively coupled plasma mass spectrometry (ICP-MS). The Au concentrations in the tumors were ~1456 and 216 ng/g-tumor in mice treated with GSH- and BSA-Au$_{25}$ NCs, respectively. These values were consistent with the biodistribution data determined from the fluorescence intensity (Figure 3b). It is worth mentioning that the Au NCs used in this study showed improved tumor accumulation relative to that of larger Au NPs.[19] For example, 6.63% ID/g deposition in tumor was observed for 20 nm PEG-Au NPs, but <1% ID/g in tumor was observed for 80 nm PEG-Au NPs.[4b] These values were much lower than those of our Au NC systems: 13.1% and 8.6% ID/g in tumors for GSH- and BSA-Au$_{25}$ NCs, respectively. In addition, the Au concentration in the tumors of mice treated with GSH-Au$_{25}$ NCs (~1456 ng/g-tumor) was much higher than those reported for Au NP-based systems (e.g., 100–300 ng/g-tumor for 5–50 nm PEG-Au NPs and 100–600 ng/g-tumor for 2–15 nm tiopronin-Au NPs).[2b, 7, 19b] Taken together, the data suggested that an enhanced uptake by the tumor tissues *via* the improved EPR effect was realized for GSH-Au$_{25}$ NCs due to their ultrasmall HD and biocompatible surface.

The encouraging results from the *in vitro* radiation therapy and *in vivo* biodistribution studies on GSH- and BSA-Au$_{25}$ NCs prompted an *in vivo* radiotherapy trial to further evaluate the potential of the new Au NC-based radiosensitizers for clinical use. Twenty-four male and 24 female nude mice bearing the U14 tumor with a tumor weight of ~10 mg-tumor/kg-body were chosen as our animal model. The mice were intraperitoneally injected with GSH- or BSA-Au$_{25}$ NCs to a concentration of 10 mg-Au/kg-body. After 0.5 h, the mice were irradiated under $^{137}$Cs gamma radiation of 3600 Ci at a 5 Gy dose. After 20 days, the tumor volumes and weights in the sacrificed mice were measured (**Figure 4**a). As compared with the control group (p<0.05), remarkable decreases of ~55% and ~38% in tumor volume were observed in mice treated with GSH- and BSA-Au$_{25}$ NCs, respectively. In addition, as compared with the mice treated by radiation only, the tumor volume decreased ~35% and ~10% in mice treated with GSH-Au$_{25}$ NCs *plus* radiation (p<0.05) and BSA-Au$_{25}$ NCs *plus* radiation (p<0.1), respectively. Figure 4b shows that the tumor weight decreased in mice treated with GSH- and BSA-Au$_{25}$ NCs *plus* radiation. Similarly, as compared with the control group (p<0.05), significant tumor weight decreases of 55% and 39% were observed in mice treated with GSH- and BSA-Au$_{25}$ NCs, respectively. The *in vivo* data further confirmed the strong radiation enhancement from the Au NCs for cancer radiotherapy. In addition, GSH-Au$_{25}$ NCs showed better accumulation in tumors and therefore stronger enhancement for cancer radiotherapy than the BSA-Au$_{25}$ NCs.

The promising *in vivo* radiotherapy data of the GSH-Au$_{25}$ NCs motivated us to study their *in vivo* cytotoxicity, which is pivotal to further developing this new class of radiosensitizer for clinical use. To achieve efficient cancer radiotherapy, the concentration of GSH-Au$_{25}$ NCs was determined to be 10 mg-Au/kg-body. This concentration was similar to those used in other studies on therapeutic Au NPs [2b, 19a] and was therefore chosen as our model dose. No obvious abnormal organ index and loss of the body weight were observed in mice treated with GSH-Au$_{25}$ NCs (Figure S4). **Figure 5**a shows the pathological results for the heart, liver, spleen, lung, and kidney in mice treated with GSH-Au$_{25}$ NCs, and there were no significant damages in these organs. In contrast, an obvious liver damage was observed in mice treated with BSA-Au$_{25}$ NCs (Figure 5b and 5c). This damage could be related to the long-term accumulation of BSA-Au$_{25}$ NCs in liver. Although a certain amount of the Au NCs (coated with GSH or BSA) were also accumulated in the genital system (e.g., testiculus), there were no obvious damages seen in these organs (Figure S5).

To further understand the toxicological response that caused the liver damage in the BSA-Au$_{25}$ NC system, the hematology and blood biochemistry after 20 days of injection were analyzed (Figure 5b and 5c). The hepatic-related serum chemistry, which was highly related to the liver damage and liver function alternation, was the focus of our analysis. An obvious increase of aspartate aminotransferase (AST) and a distinctive decrease of red blood cell (RBC) were observed in mice treated with BSA-Au$_{25}$ NCs (without radiation). There were no significant changes in other important indicators for liver injury, including alanine aminotransferase (ALT), albumin (ALB), and globulin (GLOB). Similar effects on the liver damage have recently been reported for the PEG-Au NP system, where the accumulation of PEG-Au NPs in liver could induce abnormal gene expression and lead to liver damage. The long retention time of BSA-Au$_{25}$ NCs in liver could be attributed to its relatively large HD (~6 nm) because particles in this HD range (>5.5 nm) were difficult to be excreted through renal clearance.[13] In contrast, GSH-Au$_{25}$ NCs had much smaller HD (~2.4 nm) and therefore had more efficient renal clearance. This was consistent with the observation that there was no visible toxicity in liver for the GSH-Au$_{25}$ NC system. The low *in vivo* toxicity of GSH-Au$_{25}$ NCs further paves its way to potential clinical applications. Studies on the long-term toxicity and more rigorous toxicological evaluation are needed to further advance GSH-Au$_{25}$ NCs as a new type of *metabolizable* and *efficient* clinical radiosensitizer for cancer radiotherapy.

## 3. Conclusion

In summary, a new type of radiosensitizer was constructed by integrating ultrasmall Au NCs (<2 nm) with biocompatible coating ligands (GSH and BSA). The new nano-constructs (GSH- and BSA-$Au_{25}$ NCs) inherit attractive features of both the Au core (strong radiotherapy enhancement from the Au atoms) and the coating shell (good biocompatibility conferred from the coating GSH or BSA). The ultrasmall $Au_{25}$ NCs with biocompatible coating surface displayed higher tumor accumulation *via* the improved EPR effect and therefore had a stronger enhancement for cancer radiotherapy than that of much larger Au NPs. The enhanced radiotherapy was due to the DNA damage caused by the photoelectric effect and Compton scattering of the $Au_{25}$ NCs. A remarkable decrease in tumor volume and weight was achieved by using the GSH-$Au_{25}$ NCs as the radiosensitizer. In addition, the hydrodynamically ultrasmall GSH-$Au_{25}$ NCs (HD ~2.4 nm) showed very efficient renal clearance and therefore had no obvious toxicity in the body, whereas the hydrodynamically larger BSA-$Au_{25}$ NCs (HD ~6 nm) could not be efficiently removed by the kidney and therefore caused the liver damage. This work is of interest not only because it presents a new type of promising radiosensitizers that have preferential accumulation in tumors, strong radiotherapy enhancement, and can be metabolized after the treatment, but also because it exemplifies a good approach to improve the biocompatibility of functional nanomaterials by simply using a naturally occurring biomolecule (e.g., GSH) as the coating ligand.

## 4. Experimental Section

*Synthesis and Characterizations of GSH- and BSA-Au$_{25}$ NCs:* The synthesis and purification of GSH- and BSA-Au$_{25}$ NCs followed published procedures.[14] In a typical synthesis of GSH-Au$_{25}$ NCs, GSH in the reduced form (40 μmol) was mixed with a methanol solution of HAuCl$_4$ (20 mL, 5 mM) at 4 °C for 30 min. An aqueous solution of NaBH$_4$ (5 mL, 0.2 M, at 0 °C) was then injected rapidly into the reaction mixture under vigorous stirring. The mixture was allowed to react at 4 °C for 1 h. The precipitate was collected and washed with methanol for three times. The precipitate was then dissolved in water (5 mL), and GSH (30 mg) was added to the solution. The solution was stirred and incubated at 55 °C for 3 h. UV-vis absorption spectroscopy was used to monitor the extent of reaction. In a typical synthesis of BSA-Au$_{25}$ NCs, an aqueous solution of HAuCl$_4$ (5 mL, 10 mM, 37 °C) was mixed with a BSA solution (5 mL, 50 mg/mL, 37 °C) under vigorous stirring. After two minutes, NaOH solution (0.5 mL, 1 M) was introduced, and the mixture was incubated at 37 °C for 12 h. Both GSH- and BSA-Au NCs were purified by using dialysis bags with a molecular weight cutoff (MWCO) of 3 kDa. The purified Au NCs were stored in fridge at 4 °C, and were ready for use.

The hydrodynamic diameter (HD) distributions of the as-synthesized Au NCs were determined by dynamic light scattering (DLS) on a NanoZS Zetasizer (Malvern). The DLS data were acquired in the phase analysis light scattering mode at 25 °C, and the sample solutions were prepared by dissolving the Au NCs in 10 mM phosphate-buffered saline (PBS) solution (pH 7.0). The core sizes of the Au NCs were analyzed by transmission electron microscopy (TEM) on a JEM-2100F (JEOL) microscope operating at 200 kV. The UV-vis absorption spectra were recorded on a UV-1800 spectrophotometer (Shimadzu). The photoluminescence (PL) spectra were measured by a F4600 fluorescence spectrophotometer (Hitachi).

*Fluorescent Labeling of GSH- and BSA-Au25 NCs:* Cy5-SE (Fanbo Biochemicals Co., Ltd. Beijing, China) was used to label $Au_{25}$ NCs according to a reported procedure.[18] In a typical modification, the as-synthesized $Au_{25}$ NCs (1.5 mL, 1 mg-Au/mL) were mixed with Cy5-SE (1.5 mL, 0.1 mg/mL), and the mixture was allowed to react for 24 h in dark. The labeled $Au_{25}$ NCs were washed several times with copious water by ultrafiltration (2 kDa MWCO membranes) until no obvious blue color was observed in the filtrate.

*In vitro Cytotoxicity Test:* Hela cells were cultured at 37 °C in humidified atmosphere with 5% $CO_2$ and low-glucose Dulbecco's modified Eagle's medium (DMEM) which contained fetal calf serum (10%), L-glutamine (2.9 mg/mL), streptomycin (1 mg/mL) and penicillin (1000 units/mL). The cells (in culture medium) were dispensed in 96-well plates (90 μL containing $10^4$ cells per well). Different concentrations of the $Au_{25}$ NCs (10 μL) were then added to each well. The effect of the concentration of $Au_{25}$ NCs was assessed using Cell Titre-Glo™ luminescent cell viability assay (Promega, Madison, WI, USA). After 24 or 48 h of treatment, 20 μL of Cell Titre-Glo™ reagent was added and mixed with the mixture in each well on an orbital shaker. The luminescence signal was recorded with a single tube luminometer (TD 20/20, Turner Biosystems Inc., Sunnyvale, CA, USA). The amount of ATP, which was proportional to the number of cells presented in culture, was determined from the assay.

*In vitro Radiation Therapy:* Hela cells ($1 \times 10^3$) were incubated in 25 $cm^2$ flasks overnight and then exposed to the $Au_{25}$ NCs (50 μg-Au/mL) for another 24 h. Cells were then irradiated under gamma-rays from $^{137}Cs$ (photon energy 662 keV) with an activity of 3600 Ci at the doses of 1, 2, 4, 6, and 8 Gy. After irradiation, cells were trypsinized, counted, and seeded in 6 cm dishes with a 5 mL culture medium. Six dishes were prepared for each dose. The cells were incubated for 10 days and then stained with crystal violet. The as-formed colonies were fixed and the surviving fraction was determined by the ratio of colony numbers in the irradiated cells to that in the untreated cells. Colonies with more than 50 cells were counted. The cell survival curve was fitted using a multi-target single-hit model ($S = 1 - (1 - e^{D/D_0})^N$), where S is the surviving fraction and D is the radiation dose. The value of $D_0$ was estimated from the fitting, and the sensitization enhancement ratio (SER) was determined by the radiation dose that led to a 50% survival of the cells.

*In vitro DNA Break:* A modified version of the alkaline COMET-assay protocol was performed to evaluate the DNA break. In a typical assay, frosted microscope slides were covered with 200 μL of 0.1% agarose in PBS. After the solidification of agarose, $2 \times 10^5$ cells suspended in 10 μL of PBS and 75 μL of 0.5% low-melting-point agarose were added to each slide. After solidification, the slides were placed in cold fresh lyses buffer [2.5 M NaCl, 100 mM disodium ethylenediaminetetraacetate (EDTA), 10 mM Tris-HCl, and 1% Triton X-100] for 1 h and subsequently in a horizontal gel electrophoresis unit (20 × 25 cm) filled with chilled electrophoresis buffer (300 mM NaOH and 1 mM $Na_2$EDTA) for 30 min. Electrophoresis was then conducted at 14 V for 1 h. The slides were drained, neutralized, and dried with ethanol after the electrophoresis. The comets were stained with ethidium bromide. The DNA damage was analyzed using Comet Assay Software Project (CASP) software that measures the tail moment.

*In vivo Imaging:* All animals were purchased, maintained, and handled using protocols approved by the Institute of Radiation Medicine, Chinese Academy of Medical Sciences (CAMS). The U14 tumor models were generated by subcutaneous injection of $2 \times 10^6$ cells suspended in 50 μL of PBS into the right shoulder of male nude mice. Before the experiments, the mice were anesthetized by chloral hydrate. The Cy5-labeled GSH-Au NCs (150 μL, 1 mg-Au/mL) and BSA-Au NCs (150 μL, 1 mg-Au/mL) were respectively intraperitoneally injected into two groups of male nude mice (three mice per group) at day 7 after the tumor inoculation when the tumor volume reached 100–120 mm$^3$. The Au NCs were then imaged using the *in vivo* fluorescence imaging system (Caliper Inc.). Visible red light with a central wavelength of 595 nm was used as the excitation source. The *in vivo* imaging wavelength range was 610–800 nm with an exposure time of 82 and 76 ms. Autofluorescence was removed using the spectral unmixing software. In the biodistribution and blood concentration measurement, the organ tissues were homogenized in the buffered formalin, and the resultant tissue suspensions were diluted 100 times. The photoluminescence intensity of the samples was measured at the excitation wavelength of 595 nm. The photoluminescence intensities of both standard samples and tissue samples were all adjusted to a linear range. The biodistribution of the Au NCs in the organs of the mice was obtained and plotted with the unit of % ID/g (percentage of the injected dose per gram-tissue).

*In vivo Radiation Therapy:* All animals were purchased, maintained, and handled using protocols approved by the Institute of Radiation Medicine, CAMS. The U14 tumor models were generated by subcutaneous injection of $2 \times 10^6$ cells suspended in 50 μL of PBS into the right shoulder of BALB/c mice. The mice were intraperitoneally treated with the GSH- and BSA-Au$_{25}$ NCs when the tumor volume reached 100–120 mm$^3$ (7 days after tumor inoculation). For each treatment, Au$_{25}$ NCs (1 mg-Au/mL) were intraperitoneally injected at a dosage of 10 mg/kg in the mice. As the control, 200 μL of saline was intraperitoneally injected into each mouse in the control group. Subsequently, the mice were irradiated by 5 Gy gamma-rays from $^{137}$Cs (photon energy 662 keV) with an activity of 3600 Ci. 48 mice were assigned to the following six groups (eight mice per group): control, GSH-Au$_{25}$ NCs, BSA-Au$_{25}$ NCs, radiation alone, GSH-Au$_{25}$ NCs + radiation, BSA-Au$_{25}$ NCs + radiation. Every group includes four male and four female mice in order to monitor the gender difference. The tumor size was measured every two or three days and calculated using the equation: tumor volume = (tumor length) × (tumor width)$^2$ / 2.

*In vivo Toxicity:* Mice were weighed and assessed for behavioral changes. At day 20 after the treatment, all mice were sacrificed, and their blood and organs were collected for hematology, biochemistry and toxicological investigation. The blood was drawn for hematology analysis (potassium EDTA collection tube) and serum biochemistry analysis (lithiumheparin collection tube) using a standard saphenous vein blood collection technique. During necropsy, liver, kidney, spleen, heart, lung, brain, genitals, tumor, and thyroid were collected and weighed. The spleen and thymus indexes ($S_x$) were used to examine the grade of changes caused by malities. The definition of $S_x$ is shown as below:

$$S_x = \frac{\text{Weight of experimental organ } (mg)}{\text{Weight of experimental animal } (g)}$$

Major organs from these mice were then fixed in 4% neutral buffered formalin, processed into paraffin, and stained with hematoxylin and eosin (H&E). Pathology was examined using a digital light microscope. The organs and original solutions of BSA- and GSH-Au$_{25}$ NCs were digested using a microwave system CEM Mars 5 (CEM, Kamp Lintfort, Germany) to determine their Au content, which was measured on an inductively coupled plasma mass spectrometer (Agilent 7500 CE, Agilent Technologies, Waldbronn, Germany).

**Supporting Information**
Supporting Information is available online from the Wiley Online Library or from the author.

**Acknowledgements**

This work is supported by the National Natural Science Foundation of China (Grant No.81000668), Natural Science Foundation of Tianjin (Grant No. 13JCQNJC13500), the Subject Development Foundation of Institute of Radiation Medicine, CAMS (Grant No.SF1207) and Foundation of Union New Star, CAMS. It is also supported with the joint lab of nanotechnology for bioapplication, which was established with Life Technologies Corp. in the National Center for Nanoscience and Technology of China. Part of this work is supported by the Ministry of Education, Singapore, under grant R-279-000-327-112.

References

[1] a) X. Huang, I. H. El-Sayed, W. Qian, M. A. El-Sayed, *J. Am. Chem. Soc.* **2006,** *128*, 2115; b) D. Kim, S. Park, J. H. Lee, Y. Y. Jeong, S. Jon, *J. Am. Chem. Soc.* **2007,** *129*, 7661; c) K. Sokolov, M. Follen, J. Aaron, I. Pavlova, A. Malpica, R. Lotan, R. Richards-Kortum, *Cancer Res.* **2003,** *63*, 1999; d) R. Elghanian, J. J. Storhoff, R. C. Mucic, R. L. Letsinger, C. A. Mirkin, *Science* **1997,** *277*, 1078; e) J. Chen, F. Saeki, J. Benjamin, H. Cang, M. J. Cobb, Z. Y. Li, L. Au, H. Zhang, M. B. Kimmey, X. Li, Y. Xia, *Nano Lett.* **2005,** *5*, 473; f) N. L. Rosi, D. A. Giljohann, C. S. Thaxton, A. K. R. Lytton-Jean, M. S. Han, C. A. Mirkin, *Science* **2006,** *312*, 1027; g) S. Wang, P. Huang, L. Nie, R. Xing, D. Liu, Z. Wang, J. Lin, S. Chen, G. Niu, G. Lu, X. Chen, *Adv. Mater.* **2013,** doi: 10.1002/adma.201204623; h) N. A. Kotov, F. Stellacci, *Adv. Mater.* **2008,** *20*, 4221; i) T. K. Sau, A. L. Rogach, F. Jäckel, T. A. Klar, J. Feldmann, *Adv. Mater.* **2010,** *22*, 1805; j) H. Cheng, C. J. Kastrup, R. Ramanathan, D. J. Siegwart, M. Ma, S. R. Bogatyrev, Q. Xu, K. A. Whitehead, R. Langer, D. G. Anderson, *ACS Nano* **2010,** *4*, 625; k) M. I. Setyawati, C. Y. Tay, S. L. Chia, S. L. Goh, W. Fang, M. J. Neo, H. C. Chong, S. M. Tan, S. C. J. Loo, K. W. Ng, J. Xie, C. N. Ong, N. S. Tan, D. T. Leong, *Nat. Commun.* **2013,** *4,* 1673.

[2] a) J. F. Hainfeld, D. N. Slatkin, H. M. Smilowitz, *Phys. Med. Biol.* **2004,** *49*, N309; b) X. D. Zhang, D. Wu, X. Shen, J. Chen, Y. M. Sun, P. X. Liu, X. J. Liang, *Biomaterials* **2012**, *33*, 6408.

[3] a) R. Jin, *Nanoscale* **2009,** *2*, 343; b) Q. Zhang, J. Xie, Y. Yu, J. Y. Lee, *Nanoscale* **2010,** *2*, 1962; c) Y. Negishi, Y. Takasugi, S. Sato, H. Yao, K. Kimura, T. Tsukuda, *J. Am. Chem. Soc.* **2004,** *126*, 6518; d) Y. Negishi, K. Nobusada, T. Tsukuda, *J. Am. Chem. Soc.* **2005,** *127*, 5261; e) S. Choi, R. M. Dickson, J. Yu, *Chem. Soc. Rev.* **2012,** *41*, 1867.

[4] a) S. D. Perrault, C. Walkey, T. Jennings, H. C. Fischer, W. C. W. Chan, *Nano Lett.* **2009,** *9*, 1909; b) G. Zhang, Z. Yang, W. Lu, R. Zhang, Q. Huang, M. Tian, L. Li, D. Liang, C.

Li, *Biomaterials* **2009,** *30*, 1928; c) L. Y. T. Chou, W. C. W. Chan, *Adv. Healthcare Mater.* **2012**. *1*, 714.

[5]     Y. J. Li, A. L. Perkins, Y. Su, Y. Ma, L. Colson, D. A. Horne, Y. Chen, *Proc. Natl. Acad. Sci. U. S. A.* **2012,** *109*, 4092.

[6]     a) Y. Wang, Y. Liu, H. Luehmann, X. Xia, P. Brown, C. Jarreau, M. Welch, Y. Xia, *ACS Nano* **2012,** *6*, 5880; b) L. Balogh, S. S. Nigavekar, B. M. Nair, W. Lesniak, C. Zhang, L. Y. Sung, M. S. T. Kariapper, A. El-Jawahri, M. Llanes, B. Bolton, F. Mamou, W. Tan, A. Hutson, L. Minc, M. K. Khan, *Nanomed:NBM.* **2007,** *3*, 281.

[7]     K. Huang, H. Ma, J. Liu, S. Huo, A. Kumar, T. Wei, X. Zhang, S. Jin, Y. Gan, P. C. Wang, S. He, X. Zhang, and X. J. Liang, *ACS Nano* **2012,** *6*, 4483.

[8]     a) A. Verma, F. Stellacci, *Small* **2009,** *6*, 12; b) A. Verma, O. Uzun, Y. Hu, Y. Hu, H. S. Han, N. Watson, S. Chen, D. J. Irvine, F. Stellacci, *Nat. Mater.* **2008,** *7*, 588.

[9]     a) C. Zhou, M. Long, Y. Qin, X. Sun, J. Zheng, *Angew. Chem. Int. Ed.* **2011,** *123*, 3226; b) C. Zhou, G. Hao, P. Thomas, J. Liu, M. Yu, S. Sun, O. K. Öz, X. Sun, J. Zheng, *Angew. Chem. Int. Ed.* **2012,** *124*, 10265.

[10]    X. D. Zhang, D. Wu, X. Shen, P. X. Liu, F. Y. Fan, S. J. Fan, *Biomaterials* **2012,** *33*, 6408.

[11]    a) J. Yu, S. Choi, R. M. Dickson, *Angew. Chem. Int. Ed.* **2009,** *121*, 324; b) J. Yu, S. A. Patel, R. M. Dickson, *Angew. Chem. Int. Ed.* **2007,** *119*, 2074; c) Y. Zhang, F. Zheng, T. Yang, W. Zhou, Y. Liu, N. Man, L. Zhang, N. Jin, Q. Dou, Y. Zhang, *Nat. Mater.* **2012,** *11*, 817.

[12]    a) W. H. De Jong, W. I. Hagens, P. Krystek, M. C. Burger, A. J. A. M. Sips, R. E. Geertsma, *Biomaterials* **2008,** *29*, 1912; b) J. Lipka, M. Semmler-Behnke, R. A. Sperling, A. Wenk, S. Takenaka, C. Schleh, T. Kissel, W. J. Parak, W. G. Kreyling, *Biomaterials* **2010,** *31*, 6574.

[13]     H. S. Choi, W. Liu, P. Misra, E. Tanaka, J. P. Zimmer, B. I. Ipe, M. G. Bawendi, J. V. Frangioni, *Nat. Biotech.* **2007,** *25*, 1165.

[14]     a) J. Xie, Y. Zheng, J. Y. Ying, *J. Am. Chem. Soc.* **2009,** *131*, 888; b) Y. Shichibu, Y. Negishi, H. Tsunoyama, M. Kanehara, T. Teranishi, T. Tsukuda, *Small* **2007,** *3*, 835.

[15]     M. Zhu, C. M. Aikens, F. J. Hollander, G. C. Schatz, R. Jin, *J. Am. Chem. Soc.* **2008,** *130*, 5883.

[16]     Y. Shichibu, Y. Negishi, T. Tsukuda, T. Teranishi, *J. Am. Chem. Soc.* **2005,** *127*, 13464.

[17]     Z. Wu, R. Jin, *Nano Lett.* **2010,** *10*, 2568.

[18]     L. Cheng, K. Yang, Q. Chen, Z. Liu, *ACS Nano* **2012,** *6*, 5605.

[19]     a) G. von Maltzahn, J. H. Park, A. Agrawal, N. K. Bandaru, S. K. Das, M. J. Sailor, S. N. Bhatia, *Cancer Res.* **2009,** *69*, 3892; b) X. Huang, X. Peng, Y. Wang, D. M. Shin, M. A. El-Sayed, S. Nie, *ACS Nano* **2010,** *4*, 5887.

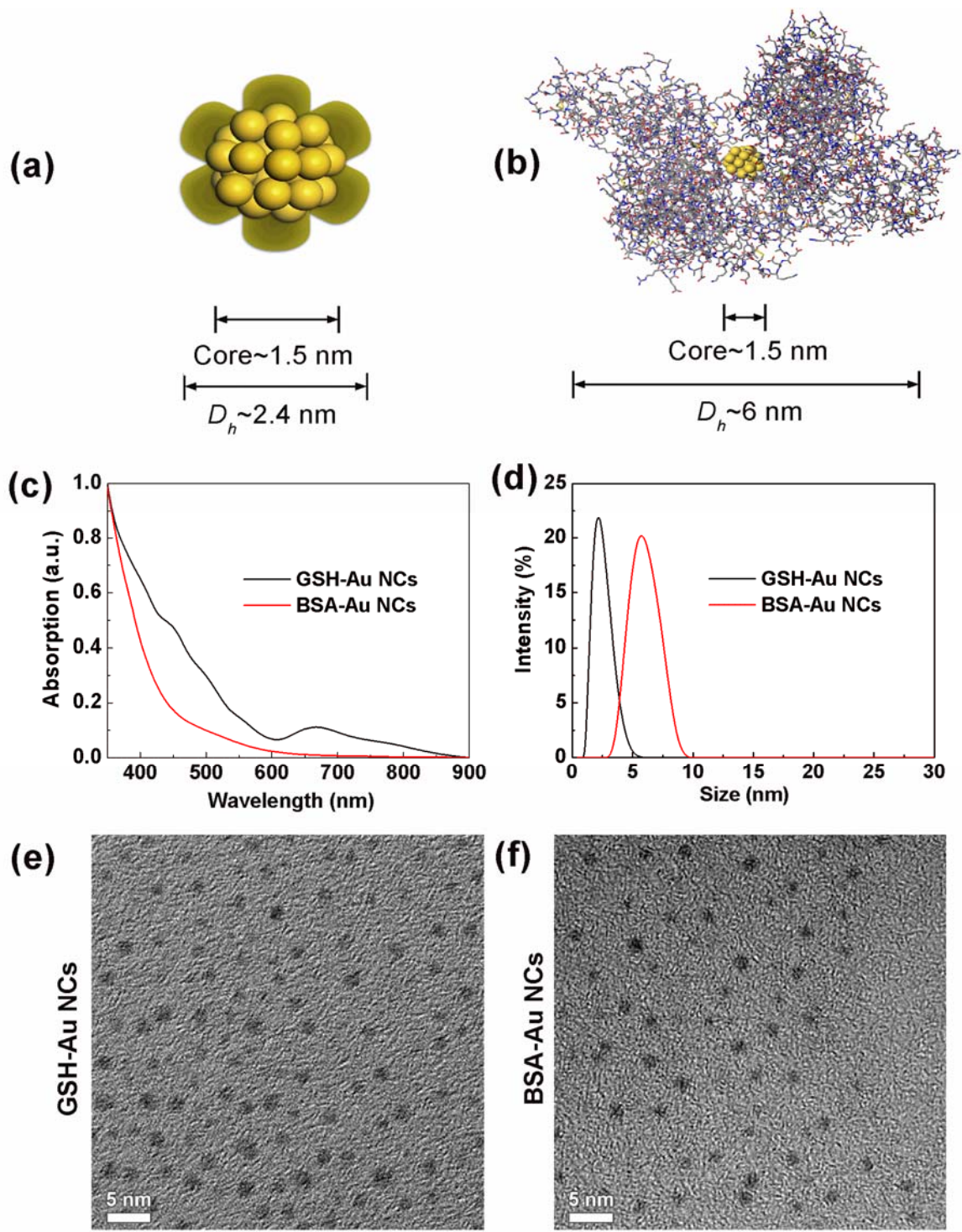

**Figure 1.** Schematic illustration of the core-shell structure of (a) GSH-Au$_{25}$ NCs and (b) BSA-Au$_{25}$ NCs. (c) UV-vis and (d) DLS spectra of the as-prepared GSH-Au$_{25}$ NCs (black line) and BSA-Au$_{25}$ NCs (red line). Representative TEM images of the as-prepared (e) GSH-Au$_{25}$ NCs and (f) BSA-Au$_{25}$ NCs.

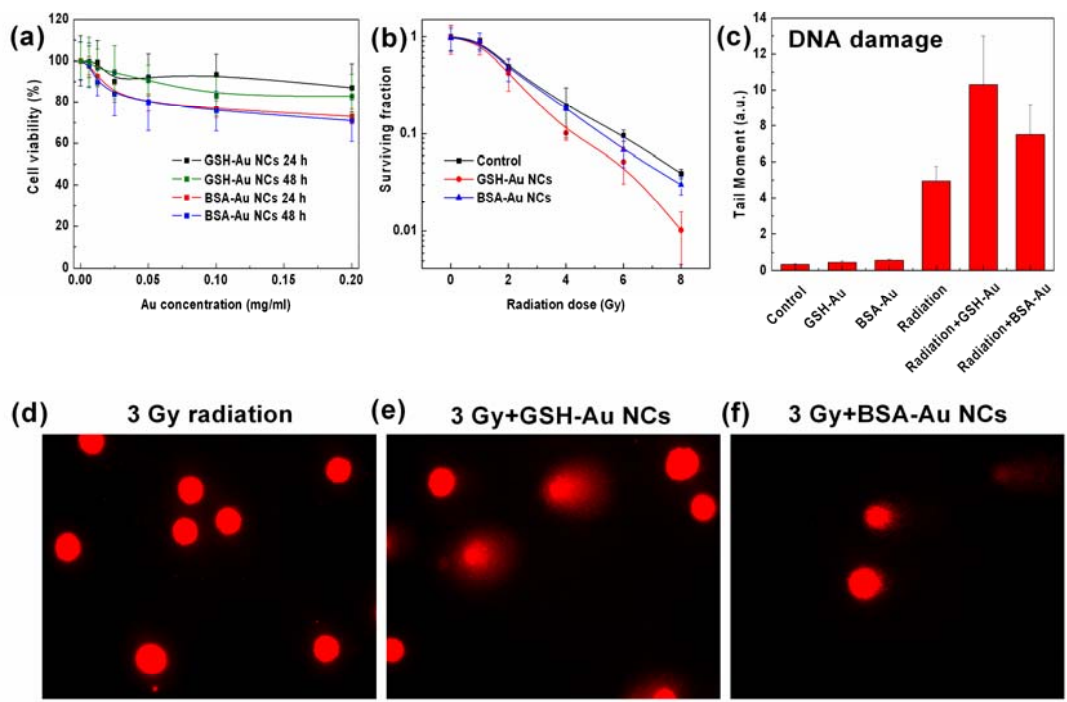

**Figure 2.** (a) Viability of Hela cells after incubation with GSH- and BSA-Au$_{25}$ NCs for 24 and 48 h. (b) Viability of Hela cells treated with only radiation (control, black line), GSH-Au$_{25}$ NCs (50 μg-Au/mL) + radiation (red line), and BSA-Au$_{25}$ NCs (50 μg-Au/mL) + radiation (blue line). (c) Tail moment of Hela cells treated with GSH-Au$_{25}$ NCs, BSA-Au$_{25}$ NCs, radiation (3 Gy), GSH-Au$_{25}$ NCs + radiation (3 Gy), and BSA-Au$_{25}$ NCs + radiation (3 Gy). Representative cell images of fluorescent DNA stain of (d) radiation group, (e) GSH-Au$_{25}$ NCs + radiation (3 Gy), and (f) BSA-Au$_{25}$ NCs + radiation (3 Gy).

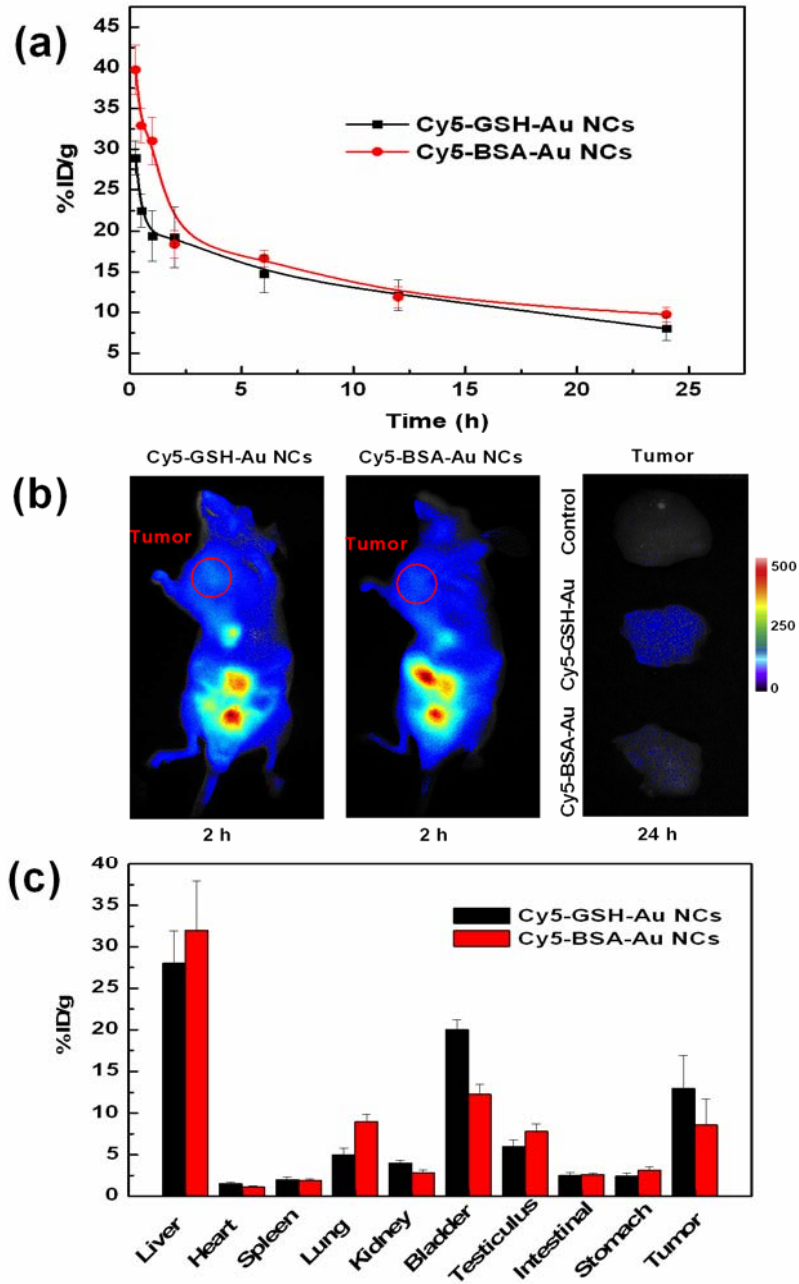

**Figure 3.** (a) *In vivo* pharmacokinetic studies of GSH-Au$_{25}$ NCs (black line) and BSA-Au$_{25}$ NCs (red line). (b) Fluorescence images of mice treated with GSH-Au$_{25}$ NCs (left panel) and BSA-Au$_{25}$ NCs (middle panel) at 2 h after injection; the right panel is the tumor images (false color) at 24 h after injection of the control group (top), GSH-Au$_{25}$ NCs group (middle), and BSA-Au$_{25}$ NCs group (down). (c) Biodistribution of GSH-Au$_{25}$ NCs (black column) and BSA-Au$_{25}$ NCs (red column) at 24 h after injection.

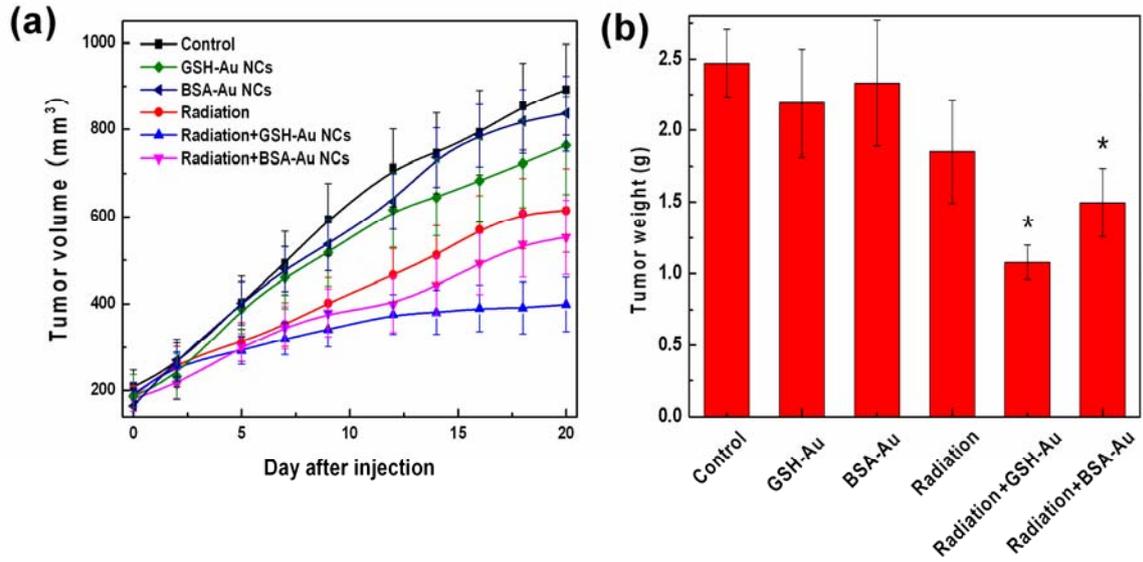

**Figure 4.** Time-course studies of tumor (a) volumes and (b) weights of mice treated with GSH- and BSA-Au$_{25}$ NCs at the concentration of 10 mg-Au/kg-body. Data was analyzed by Student's t-test and * in (b) indicates $p < 0.05$.

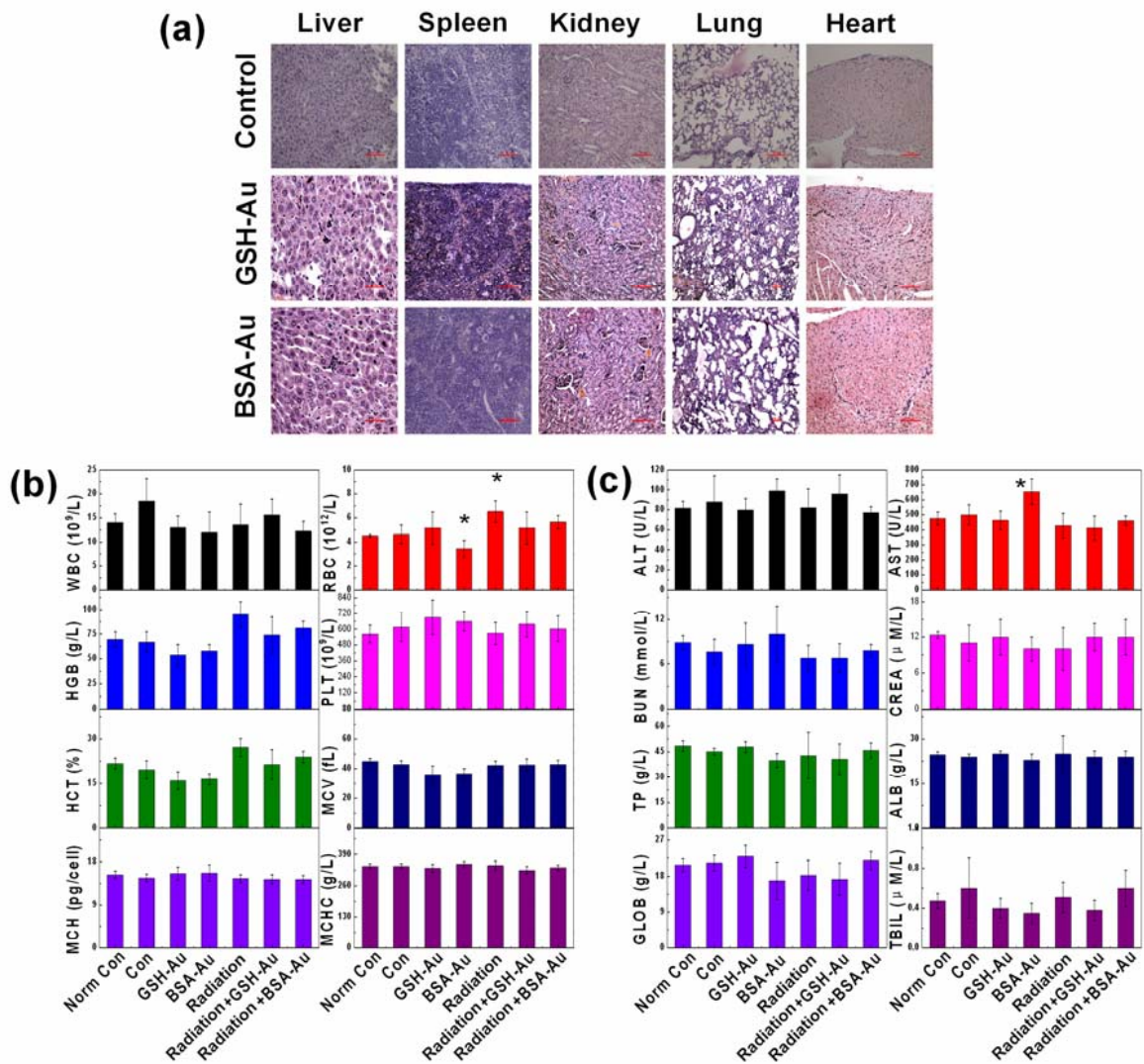

**Figure 5.** (a) Pathological data from the liver, spleen, kidney, lung, and heart of mice treated with GSH- and BSA-$Au_{25}$ NCs at the concentration of 10 mg-Au/kg-body. (b) Hematology data of mice treated with GSH- and BSA-$Au_{25}$ NCs at day 20 after injection. The results show mean and standard deviation of white blood cells (WBC), RBC, hematocrit (HCT), mean corpuscular volume (MCV), hemoglobin (HGB), platelets (PLT), mean corpuscular hemoglobin (MCH), and mean corpuscular hemoglobin concentration (MCHC). (c) Blood biochemistry analysis of mice treated with GSH- and BSA-$Au_{25}$ NCs at day 20 after injection. The results show mean and standard deviation of ALT, AST, total protein (TP), ALB, blood urea nitrogen (BUN), creatinine (CREA), GOLB, and total bilirubin (TB). Data was analyzed by Student's t-test and * in (b) and (c) indicates $p < 0.05$.

Supporting Information

**Enhanced Tumor Accumulation of Sub-2 nm Gold Nanoclusters for Cancer Radiation Therapy**

*X. Zhang, J. Chen, Z. Luo, D. Wu, X. Shen, S. Song, Y. Sun, P. Liu, J. Zhao, S. Huo, S. Fan, F.Fan, X. Liang\*, and J. Xie \**

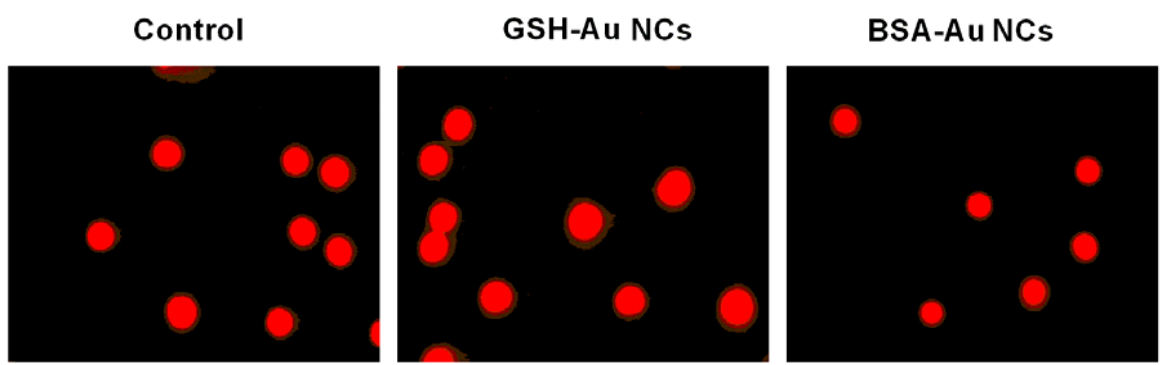

**Figure S1.** Representative images of fluorescent DNA stain of (a) untreated cells (control group) and cells treated by (b) GSH-Au$_{25}$ NCs and (c) BSA-Au$_{25}$ NCs (50 μg-Au/mL).

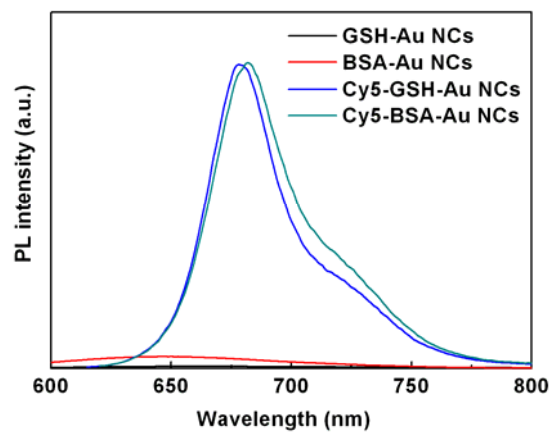

**Figure S2.** Photoemission spectra of Au$_{25}$ NCs with and without Cy5 labeling.

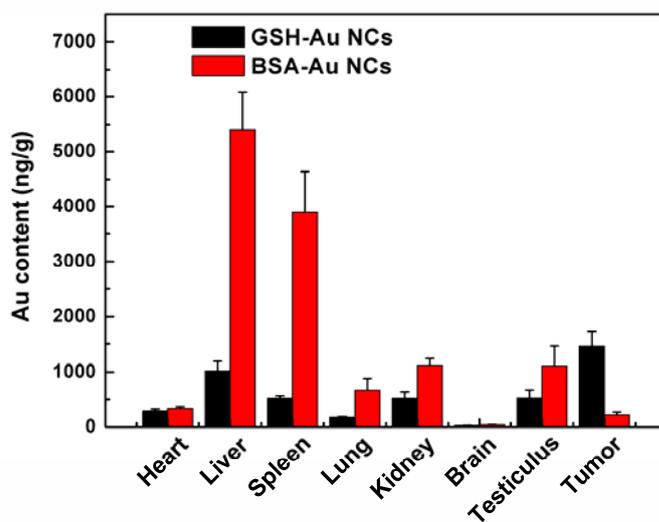

**Figure S3.** The biodistribution (in terms of Au content, determined by ICP-MS) of GSH- and BSA-Au$_{25}$ NCs at day 20 after the injection of the NCs.

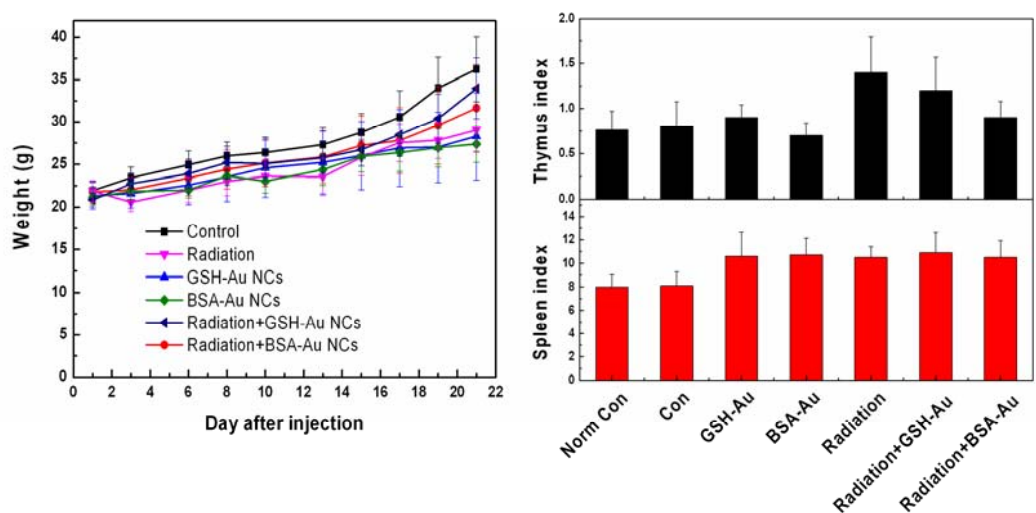

**Figure S4.** Body and organ index of the GSH- and BSA-Au$_{25}$ NCs.

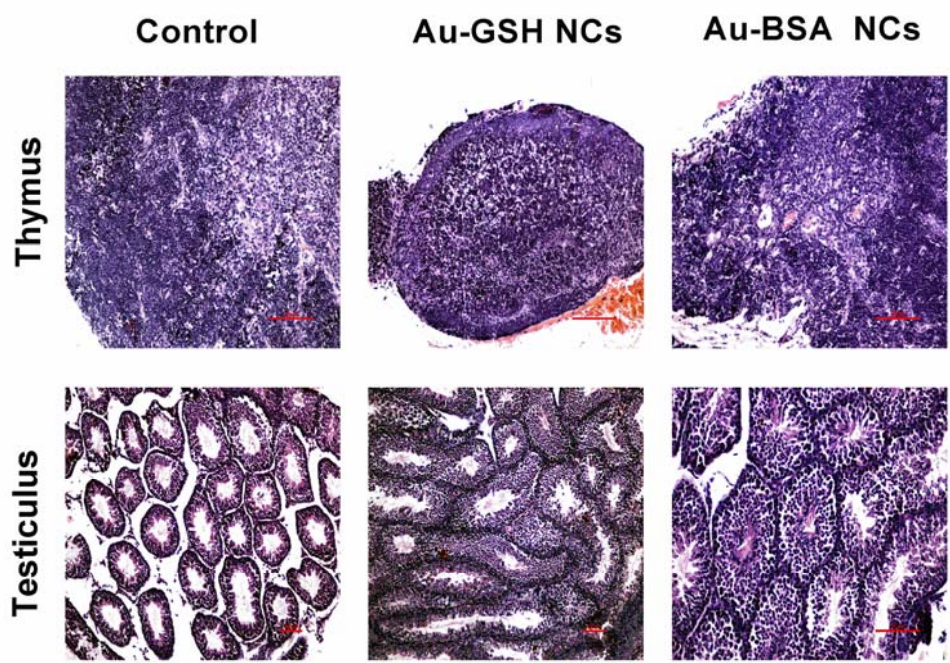

**Figure S5.** Pathological image of the thymus and testiculus from the untreated mice (control group) and the mice treated with GSH- and BSA-Au$_{25}$ NCs**.**